\input harvmac
\input epsf

\noblackbox


\def\bfone{\relax{\rm 1\kern-.35em 1}}
\def\inbar{\vrule height1.5ex width.4pt depth0pt}

\def\IC{\relax\,\hbox{$\inbar\kern-.3em{\rm C}$}}
\def\ID{\relax{\rm I\kern-.18em D}}
\def\IF{\relax{\rm I\kern-.18em F}}
\def\IH{\relax{\rm I\kern-.18em H}}
\def\II{\relax{\rm I\kern-.17em I}}
\def\IN{\relax{\rm I\kern-.18em N}}
\def\IP{\relax{\rm I\kern-.18em P}}
\def\IQ{\relax\,\hbox{$\inbar\kern-.3em{\rm Q}$}}
\def\us#1{\underline{#1}}
\def\IR{\relax{\rm I\kern-.18em R}}
\font\cmss=cmss10 \font\cmsss=cmss10 at 7pt
\def\ZZ{\relax\ifmmode\mathchoice
{\hbox{\cmss Z\kern-.4em Z}}{\hbox{\cmss Z\kern-.4em Z}}
{\lower.9pt\hbox{\cmsss Z\kern-.4em Z}}
{\lower1.2pt\hbox{\cmsss Z\kern-.4em Z}}\else{\cmss Z\kern-.4em
Z}\fi}
\def\a{\alpha} \def\b{\beta} \def\d{\delta}
 \def\g{\gamma}
 \def\la{\lambda}
 \def\s{\sigma}

\def\nup#1({Nucl.\ Phys.\ $\us {B#1}$\ (}
\def\plt#1({Phys.\ Lett.\ $\us  {B#1}$\ (}
\def\cmp#1({Comm.\ Math.\ Phys.\ $\us  {#1}$\ (}
\def\prp#1({Phys.\ Rep.\ $\us  {#1}$\ (}
\def\prl#1({Phys.\ Rev.\ Lett.\ $\us  {#1}$\ (}
\def\prv#1({Phys.\ Rev.\ $\us  {#1}$\ (}
\def\mpl#1({Mod.\ Phys.\ Let.\ $\us  {A#1}$\ (}
\def\ijmp#1({Int.\ J.\ Mod.\ Phys.\ $\us{A#1}$\ (}
\def\jag#1({Jour.\ Alg.\ Geom.\ $\us {#1}$\ (}
\def\tit#1|{{\it #1},\ }

\def\Coe#1.#2.{{#1\over #2}}

\def\coe#1.#2.{\relax{\textstyle {#1 \over #2}}\displaystyle}
\def\half{{1 \over 2}}

%
\lref\JMalda{J. Maldacena, {\it Adv. Theor. Math. Phys.} {\bf 2} (1998) 231, 
hep-th/9711200.}
\lref\IMSY{}
\lref\MDJR{M.J.~Duff and J.~Rahmfeld  \nup{481} (1996) 332, 
hep-th/9605085.}
\lref\KStelle{K.S.~Stelle  {\it BPS Branes in Supergravity,} 
in Proceedings of the ICTP Summer School in High-energy Physics 
and Cosmology, Trieste, Italy, 10 Jun -- 26 Jul 1996 and 2 Jun -- 
11 Jul 1997, hep-th/9803116.}
\lref\JMaldb{J. Maldacena, {\it Phys. Rev. Lett.} {\bf 80} (1998) 4859, 
hep-th/9803002.}
\lref\RY{S.-J. Rey and J.~Yee, {\it Macroscopic strings as heavy quarks in 
large {N} gauge theory and anti-de Sitter supergravity,} hep-th/9803001.}
\lref\RTY{S.-J. Rey, S.~Theisen and J.~Yee, {\it Nucl. Phys.} {\bf B527}
 (1998) 171, hep-th/9803135.}
\lref\BISY{A.~Brandhuber, N.~Itzhaki, J.~Sonnenschein and S.~Yankielowicz,
{\it Wilson Loops in the Large $N$ Limit at Finite Temperature}, 
hep-th/9803137}
\lref\BISYII{A.~Brandhuber, N.~Itzhaki, J.~Sonnenschein and S.~Yankielowicz,
{\it Phys. Lett.} {\bf B434} (1998) 36, hep-th/9803263.}
\lref\ML{M.~Li, {\it JHEP} {\bf 9807} (1998) 003, hep-th/9803252;
 {\it JHEP} {\bf 9808} (1998) 014,
hep-th/9804175.}
\lref\DP{ U.~Danielsson and  A.~Polychronakos, {\it Quarks, monopoles and 
dyons at large $N$,} hep-th/9804141.}
\lref\JMin{J.~Minahan, {\it Quark-Monopole Potentials in Large $N$ Super 
Yang-Mills}, hep-th/9803111.}
\lref\LanLif{L.D.~Landau and E.M.~Lifshitz, {\it Mechanics,}
Third Edition, Pergamon Press, (1973).}
\lref\Wittherm{E.~Witten, , {Adv. Theor. Math. Phys.} {\bf 2} (1998) 505,
hep-th/9803131.}
\lref\MW{J. Minahan and N. Warner}
\lref\Arvis{J.~Arvis, {\it Phys. Lett.} {\bf 127B} (1983) 106.}
\lref\Alv{O.~Alvarez, {\it Phys. Rev. D} {\bf 24} (1981) 440.}
\lref\Luscher{M.~L\"uscher, {\it Nucl. Phys.} {\bf B180} (1981) 317.}
\lref\HashOz{A.~Hashimoto and Y.~Oz, {\it Aspects of QCD Dynamics from String 
Theory}, hep-th/9809106.}
\lref\KJMN{R.de Mello Koch, A.Jevicki, M.Mihailescu and J.P.Nunes, 
Phys.Rev. {\bf D58} (1998) 105009, hep-th/9806125}
\lref\KT{I. Klebanov and A. Tseytlin, {\it D-Branes and Dual Gauge Theories 
in Type 0 Strings}, hep-th/9811035.}
\lref\BG{O. Bergman and M. Gaberdiel, {\it Nucl. Phys.} {\bf B499} (1997) 183,
hep-th/9701137.}
\lref\Horava{P. Horava, {\it On QCD String Theory and AdS Dynamics},
hep-th/9811028.}
\lref\Polyakov{A. Polyakov, {\it Nucl. Phys.} {\bf B268} (1986) 406.}
\lref\Zyskin{M. Zyskin, {\it Phys. Lett.} {\bf B439} (1998) 373, 
hep-th/9806128.}
\lref\PP{A. Peet, J. Polchinski, {\it UV/IR Relations in AdS Dynamics},
hep-th/9809022.}
\lref\ABKS{O. Aharony, M. Berkooz, D. Kutasov and N. Seiberg,
{\it Linear Dilatons, NS5-branes and Holography}, hep-th/9808149.}
\lref\GO{D. Gross and H. Ooguri, {\it Phys. Rev.} {\bf D58} (1998) 106002,
hep-th/9805129.}
\lref\COOT{C. Csaki, H. Ooguri, Y. Oz and J. Terning, {\it Glueball Mass 
Spectrum From Supergravity}, hep-th/9806021.}
\lref\CORT{C. Csaki, Y. Oz, J. Russo and J. Terning, {\it Large $N$ QCD from 
Rotating Branes}, hep-th/9810186.}
\lref\Russo{J. Russo, {\it  New Compactifications of Supergravities and Large 
$N$ QCD},  hep-th/9808117.}
\lref\GKP{S. Gubser, I. Klebanov and  A. Polyakov, {\it Phys. Lett.} 
{\bf B428} (1998) 105, hep-th/9802109.}
\lref\Witten{E. Witten, {\it Adv. Theor. Math. Phys.} {\bf 2} (1998) 253,
hep-th/9802150.}
\lref\DH{L. Dixon and J. Harvey, {\it Nucl. Phys.} {\bf B274} (1986) 93.}
\lref\SW{N. Seiberg and E. Witten, {\it Nucl. Phys.} {\bf B276} (1986) 272.}
\lref\Gross{D. Gross, Strings '98 talk.}
\lref\GKPeet{S. Gubser, I. Klebanov and A. Peet, {\it Phys. Rev. D} {\bf 54}
 (1996) 3915, hep-th/9602135.}
\lref\Cave{A. Polyakov, {\it The Wall of the Cave}, hep-th/9809057}
\lref\JM{J. A. Minahan, {\it Glueball Mass Spectra and Other Issues for 
Supergravity Duals of QCD Models}, hep-th/9811156.}
\lref\KTII{I. Klebanov and A. Tseytlin, {\it Asymptotic Freedom and Infrared 
Behavior in the Type 0 String Approach to Gauge Theory}, hep-th/9812089.}
\lref\KTIII{I. Klebanov and A. Tseytlin, {\it A Non-supersymmetric Large $N$ 
CFT from Type 0 String Theory}, hep-th/9901101.}
\lref\GrOl{J. Greensite and P. Olesen, {\it Remarks on the Heavy Quark 
Potential in the Supergravity Approach}, {\it JHEP} {\bf 9808} (1998) 009,
hep-th/9806235.}
\lref\ORT{H. Ooguri, H. Robins and J. Tannenhauser,  {\it Phys. Lett.} 
{\bf B437} (1998) 77, hep-th/9806171.}
\lref\Teper{M. Teper, {\it Glueball masses and other physical properties of 
$SU(N)$ gauge theories in $D=3+1$: a review of lattice results
       for theorists}, hep-th/9812187.}
\lref\MP{C. Morningstar and M. Peardon, {\it The glueball spectrum from an 
anisotropic lattice study}, hep-lat/9901004.}
\lref\GKT{S. Gubser, I. Klebanov and A. Tseytlin, {\it Coupling Constant 
Dependence in the Thermodynamics of $N=4$ Supersymmetric Yang-Mills Theory},
{\it Nucl. Phys.} {\bf B534} (1998) 202, hep-th/9805156}
\lref\MT{R. Metsaev and A. Tseytlin, {\it Type IIB superstring action in 
$AdS_5\times S^5$ background}, {\it Nucl. Phys.} {\bf B533} (1998) 109,
hep-th/9805028.}
\lref\KR{R. Kallosh and A. Rajaraman, {\it Vacua of M-theory and string 
theory}, {\it Phys. Rev. D} {\bf 58} (1998) 125003, hep-th/9805041.}
\lref\Zarembo{K. Zarembo, {\it Coleman-Weinberg Mechanism and Interaction of 
D3-Branes in Type 0 String Theory}, hep-th/990110.}
\lref\RS{J. Russo and K. Sfetsos, {\it Rotating D3 branes and QCD in three 
dimensions}, hep-th/9901056.}
\lref\KS{S. Kachru and E. Silverstein, {\it Phys. Rev. Lett.} {\bf 80}
 (1998) 4855, hep-th/9802183}
\lref\KW{I. Klebanov and E. Witten, {\it Nucl. Phys.} {\bf B536} (1998) 199,
hep-th/9807080.}
\lref\FS{A. Fayyazuddin and M. Spalinski, {\it Nucl. Phys.} {\bf B535} (1998)
 219,  hep-th/9805096.}
\lref\AFM{O. Aharony, A. Fayyazuddin and J. Maldacena, {\it JHEP} {\bf 9807}
 (1998) 013, hep-th/9806159.}
\lref\LNV{ A. Lawrence, N. Nekrasov and C. Vafa, {\it Nucl. Phys.} {\bf B533}
 (1998) 199, hep-th/9803015.}
\lref\BKV{ M. Bershadsky, Z. Kakushadze and C. Vafa, {\it Nucl. Phys.} 
{\bf B523} (1998) 59, hep-th/9803076.}
\lref\KPW{A. Khavaev, K. Pilch and  N. Warner, {\it New Vacua of Gauged $N=8$
 Supergravity}, hep-th/9812035.}
\lref\KLM{A. Karch, D. Lust and A. Miemiec, {\it New $N=1$ Superconformal 
Field Theories and their Supergravity Description}, hep-th/9901041.} 
\lref\BZ{T. Banks and A. Zaks, {\it Nucl.Phys.} {\bf B196} (1982) 189.}
\lref\AG{E. Alvarez and C. Gomez, {\it Non-Critical Confining Strings and the 
Renormalization Group}, hep-th/9902012.} 
\lref\BaGr{T. Banks and M. Green, {\it JHEP} {\bf 9805} (1998) 002, 
hep-th/9804170.}
\lref\CRST{C. Csaki, J. Russo, K. Sfetsos and J. Terning, {\it
Supergravity Models for $3+1$ Dimensional QCD}, hep-th/9902067.}
\lref\Gubser{S. Gubser, {\it Phys. Rev. D} {\bf 59} (1999) 025006, 
hep-th/9807164.}
\lref\DZ{J. Distler and F. Zamora,{\it Non-Supersymmetric Conformal Field 
Theories from Stable Anti-de Sitter Spaces}, hep-th/9810206.}
\lref\BF{P. Breitenlohner and D. Freedman, {\it Phys. Lett.} {\bf 115B} 
(1982) 197; {\it Ann. Phys.} {\bf 144} (1982) 197.}
\lref\MeTo{L. Mezincescu and P. Townsend, {\it Ann. Phys.} {\bf 160} (1985) 
406.}
\lref\Gar{M. Garousi, {\it String Scattering from D-branes in Type 0 Theories},
hep-th/9901085.}
       
%
%
\Title{\vbox{
\hbox{IASSNS-HEP-99/14}
\hbox{\tt hep-th/9902074}
}}{\vbox{\centerline{\hbox{Asymptotic Freedom and Confinement from 
}}
\vskip 8 pt
\centerline{ \hbox{Type 0 String Theory}}}}
\centerline{Joseph A. Minahan}
\bigskip
\centerline{{\it School of Natural Sciences, Institute for Advanced
Study}}
\centerline{{\it Olden Lane,
Princeton, NJ 08540, USA}}
\bigskip
We argue that there are generic solutions to the type 0 gravity
equations of motion that are confining in the infrared and have log
scaling in the ultraviolet.    
The background
curvature generically diverges in the {\it IR}.  
Nevertheless, there exist solutions
where higher order string corrections appear to be exponentially 
suppressed in the {\it IR} with respect to the leading type 0 gravity terms. 
For these solutions the  tachyon flows to a fixed value.
 We show that the generic solutions lead to
a long range linear quark potential, magnetic screening and a discrete
glueball spectrum.  We also estimate some WKB glueball mass ratios and compare 
them to ratios found using finite temperature models and   lattice 
computations.

\vskip .3in

\Date{\sl {February, 1999}}
\vfil
\eject
\parskip=4pt plus 15pt minus 1pt
\baselineskip=15pt plus 2pt minus 1pt
%
\newsec{Introduction}

One of the many interesting developments to arise out of Maldacena's 
conjecture \refs{\JMalda\GKP{--}\Witten} concerns the study of large $N$ QCD.  
While it is not
known how to construct the  dual gravity theory for $SU(3)$ gauge theory
with six quark flavors, nor even the dual for any pure $SU(N)$ theory,
it is hoped that the gravity duals that are known will share
many of the properties of  everyday QCD.  The main properties
we seek are asymptotic freedom in the ultraviolet
 and confinement in the infrared.

In four space-time dimensions, the Maldacena conjecture was originally 
applied to ${\cal N}=4$ supersymmetric
Yang-Mills \JMalda\ and then to other theories that have less supersymmetry but
are still conformal \refs{\KS\LNV\BKV\FS\AFM\KW\Gubser\DZ\KPW\KLM{--}\KTIII}.  
Since the $\beta$-function is zero for these
theories, they do not behave like ordinary QCD.   

Witten has proposed supersymmetric Yang-Mills at finite temperature
as a model more suited for comparison to
 QCD \Wittherm.  A $d$ dimensional euclidean gauge theory 
at finite temperature 
 is equivalent to a theory with $d-1$ noncompact directions and a 
Euclidean time compactified on a circle.  At large
distances the theory acts like a nonsupersymmetric euclidean Yang-Mills theory
in $d-1$ dimensions.  Hence, one expects to find area law behavior and
a gap.  The gravity dual for finite temperature super Yang-Mills
is a nonextremal black hole in AdS space.  Using this, Witten was
able to argue that the bulk dilaton wave equation has a discrete
spectrum, implying a gap for the boundary theory.  He also demonstrated
 that there is
an area law for the space-like Wilson lines. It was later shown that
the finite temperature theory exhibits magnetic screening \refs{\GO,\ML}.
There has also been a mini-industry comparing  dual gravity results
to lattice results \Teper, with some
reasonable agreement \refs{\COOT\KJMN\Zyskin\HashOz\Russo\CORT\JM\RS{--}\CRST}.

However, in the {\it UV} the finite temperature
theories effectively
become  supersymmetric   with
 one extra dimension.  Thus, the
coupling does not run, but remains a free parameter.  
Related to this problem is that the QCD
scale is set by the temperature instead of by dimensional transmutation.   
So  
the glueball mass scale is roughly the temperature $T$, while   the 
string tension, with
the free parameter, is $g^2N T^2$.   In a QCD theory where the scale is
set by dimensional transmutation, one expects the glueball mass scale
to be directly related to the string scale.  

We thus seek another gravity dual to better describe QCD.  One
such candidate is the type 0 model proposed by Klebanov and Tseytlin \KT.
Ironically, this model was considered a toy model when first developed
\refs{\DH,\SW}.  Following an original suggestion by Polyakov \Cave,
Klebanov and Tseytlin argued that  one could construct the 
dual of an $SU(N)$ gauge theory with 6 real adjoint scalars by stacking
$N$ electric D3 branes of the type 0 model on top of each other.  The type
0 model has a closed string tachyon that lives in the bulk, but no open 
string tachyon that
lives on the branes.  The closed string tachyon couples to the five form
field strength, which then drives the tachyon to a nonzero expectation
value.  

In \JM\ it was shown using the equations of motion derived in \KT\ that 
the effective coupling between two heavy quarks has the desired
logarithmic fall off in the {\it UV}
\eqn\uvcoup{{g_{YM}}^2\sim {1\over \log\mu},}
where $\mu$ is the energy scale.  A possible {\it IR} solution was given in
\JM, but after a closer inspection, one finds that for generic tachyon
potentials, the {\it UV} solution does not connect to this particular {\it IR} 
solution.
In \KTII\ a different {\it IR} solution was given which corresponds to an
{\it IR} fixed point with infinite coupling and no confinement.  For a
given class of tachyon potentials, such a fixed point could conceivably
connect to the {\it UV} point in \JM. 

At first glance, 
the presence of this {\it IR} fixed point seems  sensible  since, as was
shown in \KTII, the two loop contribution to the $\b$-function
is positive.  Hence one might
expect a Banks-Zaks fixed point \BZ.  Presumably, quantum corrections can
push the value of this fixed point to infinite coupling.

On the other hand, one could argue that quantum corrections lift the masses
of the adjoint scalars. Hence this theory should behave much like pure
$SU(N)$ gauge theory, with some extra massive states.  Using this
reasoning, one should expect the type 0 theory to confine.  


In fact, using only some modest assumptions about the tachyon potentials,
we will show that the classical dual gravity equations of
motion have generic solutions in the {\it IR} which exhibit 
i) a linear quark-antiquark potential
ii) magnetic screening  and iii) a discrete glueball spectrum.  The 
glueball mass scale is related to the string scale by $m\sim\sqrt{\s}/N^{1/4}$,
with no dependence on a free parameter.
If one assumes that the tachyon reaches a fixed value in the {\it IR}, 
then there
exists  {\it IR} solutions where the curvature diverges, but where many,
if not all, $\a'$ 
corrections fall off exponentially. 

In section 2 we review the type 0 model and discuss a {\it UV} solution
with a running coupling constant.  We assume here that the
tachyon potentials satisfy a minimum number of properties, but are
otherwise generic.  In section 3 we discuss various
 {\it IR} solutions.  We argue that in order to reach the {\it IR}
solutions discussed previously in the literature requires plenty
 of fine tuning,
and even then, it is not  guaranteed that the {\it UV} solution
of section 2 can match to these solutions.  We then discuss a generic
class of {\it IR} solutions and argue that  the {\it UV}
solution is adjustable enough to match over the range of these 
{\it IR}  solutions.  In section 4 we examine the validity
of these solutions beyond the classical limit, where in general we find
that string corrections  swamp the leading order term.  However,
we find some solutions where many $\a'$ corrections are 
exponentially suppressed.  In the {\it IR} these solutions 
approach a conformal transformation
of the product space $R_+\times M_4\times S_5$, with the tachyon flowing
to a constant value.
This leads us to speculate that there is an exact solution to the full
$\s$-model whose asymptotic behavior matches the asymptotic behavior of
these classical solutions.  
 In section 5 we discuss the physical implications
of the generic {\it IR} solutions in section 3.  Given one inequality, which is
satisfied by the special solutions found in section 4, we argue
that two heavy quarks have a linear potential at large distances.  We further
argue that 
there is magnetic screening and a discrete spectrum.  
We also calculate some glueball
mass ratios using the WKB method.  The results differ somewhat from
the finite temperature results and  the errors are larger due to 
the complexity of the equations.  For the low lying masses we can make
reasonable estimates for lower bounds of mass ratios which 
 are consistent with lattice computations.  
In section 6 we present our conclusions.

\newsec{The Type $0$ Nonsupersymmetric Model.}


The type $0$ model \refs{\DH,\SW} has a closed string tachyon,
no fermions and a doubled set of
 R-R fields \refs{\DH,\SW}, and thus a doubled set of D branes \BG.  
Because of this doubling of
the  R-R fields,
one can relax the self dual constraint on the 5-form field strength.
 Hence one can have D3 branes that are electric instead of
dyonic.   The low energy
world-volume action for $N$ parallel electric D3 branes  is  $SU(N)$ QCD with 
six real adjoint
scalar fields, but no fermions.  
Hence, there is no supersymmetry and the
coupling will run.  There is no open string tachyon \BG, so
there is no tachyon in this QCD model.

Klebanov and Tseytlin argued that the massed squared of the
{\it closed} string tachyon  gets a positive shift
from the background 5 form flux \KT.  With a background flux the tachyon 
potential is not symmetric
under $T\to -T$, so a large  flux can drive
the tachyon expectation value  to a nonzero
value that is basically independent of the tachyon bare mass term and
its quartic coupling.  The tachyon field is 
a source for the dilaton, thus unlike the ${\cal N}=4$ case, the
 dilaton expectation is not constant.  

One then
makes the following ansatz for the metric \KT
\eqn\typezm{
ds^2~=~e^{\half\phi}\left(e^{\half\xi-5\eta}d\rho^2~+~e^{-\half\xi}dx_{||}^2
~+~e^{\half\xi-\eta}d\Omega_5^2\right),
}
where $\phi$, $\xi$ and $\eta$ are functions of $\rho$ only.  The equations
of motion then reduce to a Toda like system with an action 
\KT
\eqn\toda{\eqalign{
S~=&~\int d\rho\left[\half \dot\phi^2+\half\dot\xi^2+{1\over4}\dot T^2
-5\dot\eta^2-
V(\phi,\xi,\eta,T)\right]\cr
V(\phi,\xi,\eta,T)~=&~g(T)e^{\half\phi+\half\xi-5\eta}+20e^{-4\eta}-
Q^2f^{-1}(T)e^{-2\xi},
}}
and a constraint
\eqn\constraint{
\half \dot\phi^2+\half\dot\xi^2+{1\over4}\dot T^2-5\dot\eta^2+
V(\phi,\xi,\eta)~=~0.
}
That the left hand side of \constraint\ is a constant
follows from the equations of
motion.  The fact that the constant is zero follows
directly from the dilaton equations of motion and the Einstein equation.
$Q$ is the total D3 brane charge which is proportional to $N$, $T$ is the
tachyon field and $f(T)$ is a function given by \KT
\eqn\fT{
f(T)~=~1+T+\half T^2.
}
This describes the tachyon coupling to the five form flux. The bare 
tachyon potential is the negative of $g(T)$
\eqn\tachpot{
g(T)~=~\half T^2~-~\la T^4
}
where we have included the quartic piece of the potential.  The coefficient
$\la$ of the quartic coupling has yet to be determined unambiguously, so for 
the
purposes of this paper, we will treat it as a parameter.  Actually, the
details of $f(T)$ and $g(T)$ are not that important, except that we
require that $f(T)$ have a minimum at some finite value of $T$ and that
$g(T)$ is positive at that value.  We will often consider the special
case where $g'(T)=0$ when $f'(T)=0$.

The equations of motion derived from the action in \toda\ are
\eqn\eqmot{\eqalign{
\ddot\xi~+~\half g(T)e^{\half\phi+\half\xi-5\eta}~+~2{Q^2\over f(T)}e^{-2\xi}
~&=~0\cr
\ddot\eta~+~\half g(T)e^{\half\phi+\half\xi-5\eta}~+~8e^{-4\eta}~&=~0\cr
\ddot\phi~+~\half g(T)e^{\half\phi+\half\xi-5\eta}~&=~0\cr
\ddot T~+~2 g'(T)e^{\half\phi+\half\xi-5\eta}~+~2{Q^2f'(T)\over f^2(T)}
e^{-2\xi}~&=~0.
}}
For large $Q$, $g(T)$ plays a secondary role to $f(T)$ and so
the tachyon expectation value is determined by setting
 $f'(T)=0$.  As a first approximation, we may assume that the tachyon is a
constant $T=T_0$. 
If $T_0=0$ then the solution reduces to the ${\cal N}=4$ solution.  
When $T$ is
nonzero, then all fields are coupled and there is no known analytic solution.
However, we can attempt to find approximate solutions that are valid in
the {\it UV} and {\it  IR} regions.  

An asymptotic solution valid in the  {\it UV} was given in \JM.
As a first approximation one can take  the dilaton field to
be relatively constant, at least compared with $\xi$ and $\eta$.  Assuming
a constant $\phi$, the equations for
$\xi$ and $\eta$ can be solved exactly, at least in the near horizon limit.
In this case we find 
\eqn\appsol{\eqalign{
&e^{\xi}~=~C_1\rho \qquad\qquad e^\eta~=~C_2\rho^{1/2}\cr
&{1\over2} g(T_0){C_1^{1/2}\over C_2^5}e^{\ha\phi}~+~{2Q^2\over C_1^2f(T_0)}
~-~1~=~0\cr
&{5}g(T_0){C_1^{1/2}\over C_2^5}e^{\ha\phi}~+~{80\over C_2^4}~-~5~=~0.
}}
One can easily check that this satisfies the constraint equation in 
\constraint. If we plug this back into the metric, we find that the solution
is still $AdS_5\times S_5$, but the curvatures
 of the two spaces no longer match;
 $S_5$ now has smaller curvature then $AdS_5$.
In this case the Ricci scalar for the total space is proportional to
\eqn\ricci{
R~\sim~ g(T_0) e^{\half\phi}.
}

Using the  $\xi$ and $\eta$ solutions as inputs, we can go back and find
an approximate solution for
 $\phi$ in terms of $\rho$. Using the ansatz
 $e^{\half\phi}=C_0(\log(\rho/\rho_0))^\a$, and
plugging this into the equation of motion for $\phi$ in \eqmot,
we find that the ansatz is a leading order solution if $\a=-1$ and
 $C_0=-4C_2^5/(g(T_0)\sqrt{C_1})$.  $\rho_0$ is an integration constant and we 
assume that $\rho_0>>1$ in order that the gauge theory length scale is
much greater than the string scale.  Setting $\rho=u^{-4}$, and using the
lowest order solutions for $C_1$ and $C_2$ from \appsol, we learn that the
leading order behavior for the coupling is
\eqn\couplo{
e^{-\phi}~=~{1\over g_{YM}^2}
~=~{Qg^2(T_0)\left(\log {u\over u_0}\right)^2\over 
1024\sqrt{2f(T_0)}}.
}
One can easily check
that to leading order in $1/\log u$, the constraint equation is still
satisfied.  We can also estimate the range of validity for this solution.
Computing the leading order corrections to $C_1$ and $C_2$, one finds that
\eqn\corrcs{
C_1~=~{2Q\over\sqrt{2f(T_0)}}\left(1+{1\over 4\log {u\over u_0}}\right)\qquad
C_2~=~2\left(1+{1\over 4\log {u\over u_0}}\right).
}
We can also compare the terms in the potential that depend on the tachyon.
Since 
\eqn\tachterms{
{1\over2}g(T_0) e^{\half\phi+\half\xi-5\eta}\sim {u^8\over 2\log {u\over u_0}}
\qquad
Q^2f^{-1}(T_0)e^{-2\xi}\sim {u^8\over2},
}
our solution with  constant $T=T_0$ and $f'(T_0)=0$ is valid
 so long as $\log(u/u_0)>>1$.  While we are treating $\la$ as a parameter,
we should note that it is crucial that $g(T)>0$ when $f'(T)=0$.  Hence
we assume that $0<\la<1/2$.  The lower bound on $\la$ is so that the complete
tachyon potential is bounded below.

The metric in the large $u$ limit is 
\eqn\metlu{
ds^2~=~{16\over g(T_0)\log {u\over u_0}}\left({du^2\over u^2}+{\sqrt{2f(T_0)}
\over2Q}
\left(1+{1\over\log {u\over u_0}}\right)u^2dx_{||}^2+
\left(1+{1\over\log {u\over u_0}}\right)
d\Omega_5^2\right).
}
Hence we can  trust the classical 
dual gravity solution only if $g(T_0)<<1$, since
 $\log (u/u_0)>>1$.  However, for generic values of $\la$,
$g(T_0)\sim1$ when $f'(T_0)=0$, hence it would seem that the  classical
 result
is not  particularly trustworthy.  As it happens, the situation
is not completely hopeless \KTII.  While the curvature for \metlu\ diverges
in the large $u$ limit, the Weyl tensor actually
falls off as 
\eqn\weyl{
|C|~\sim~{1\over\log u}.
}
This is a consequence of the conformal invariance in the large $u$ limit, where
the space-time approaches that of $AdS_5\times S_5$, with divergent
curvature.  Because of this fall off in $|C|$, the ${\a'}^3$ 
correction is of the same order as the classical term.
Hence, while quantitative results might not be exact,
the classical 
dual gravity solutions could capture the true qualitative behavior
for this $SU(N)$ gauge theory.

It would seem that the dual gravity calculation failed  its first test:
 the prediction in \couplo\ is 
 that $e^{-\phi}$ has a log squared dependence instead of 
the linear log behavior found in perturbative
Yang-Mills.  However, the physical coupling is  determined by
finding the potential between two heavy quarks.
 Using the Wilson line
computation of \refs{\RY,\JMaldb}, one finds that the quark potential is
given by \JM
\eqn\qp{
V~\approx~-~{128\ \pi^3\over  \Gamma({1\over4})^4g(T_0)L\log(L_0/L)}
\qquad\qquad L<<L_0}
where $L_0$ is some length that can be adjusted to
be much longer than the string scale.  Hence,  the effective coupling between 
a heavy quark and its antiquark
{\it does} appear with the expected log dependence. 
In deriving \qp\ we have substituted  $R^2={16\over g(T_0)\log (u/u_0)}$ into 
the quark potentials found in 
in \refs{\RY,\JMaldb}.  A  perturbative Yang-Mills computation
gives a quark potential proportional to $g^2N$.  Since the perturbative
Yang-Mills coupling behaves like $g^2N\sim 1/\log(L_0/L)$, we find that
the dual gravity computation results in an effective coupling with the
desired log dependence.  The actual coefficient is related to the
one loop $\b$-function.  Unfortunately, it is clear that the coefficient is 
model dependent  and we don't know enough to actually
compute this using the gravity dual.  Even if we did, we should not
necessarily expect to get the correct answer since stringy corrections
will be of order 1.  


As we move away from the {\it UV} point at $u=\infty$, the tachyon will begin
moving away from $T_0$.  Solving the equations of motion to the
next leading order we find that \KTII
\eqn\asymp{\eqalign{
T~&=~T_0~-4{g'(T_0)\over g(T_0)}{1\over \log\rho}~+~{\rm O}
\left({\log(-\log\rho)\over \log^2\rho}\right)\cr
\phi~&=~ -2\log(C_0\log\rho)~-~\left(7+8\left({g'(T_0)\over g(T_0)}\right)^2
\right)\left({\log(-\log\rho)\over \log\rho}\right)~+~{B\over\log\rho}\cr
&\qquad\qquad~+~{\rm O}\left({\log^2(-\log\rho)\over \log^2\rho}\right)\cr
\xi~&=~\log\left(\sqrt{2f^{-1}(T_0)}~Q\rho\right)~-~{1\over\log\rho}~+~{\rm O}
\left({\log(-\log\rho)\over \log^2\rho}\right)\cr
\eta~&=~\half\log(4\rho)~-~{1\over\log\rho}~+~{\rm O}
\left({\log(-\log\rho)\over \log^2\rho}\right).
}}
The constant $B$ in the $\phi$ expansion is an integration constant that
can be removed under a conformal rescaling. Notice that $Q$ only appears
as an overall factor in front of $e^\xi$.
We also notice that the sign of $g'(T_0)$ determines whether or not the
tachyon expectation value is driven toward or away from zero as one
moves away from $\rho=0$.  There is also the possibility that $g'(T_0)=0$,
in which case the tachyon expectation value is unchanged as the system varies.

 Another interesting point concerns the ratio of the two and one
loop contributions to the $\b$-function \KTII.  To next to leading order
the effective coupling $e^{\half\phi}$ is given by
\eqn\twoloop{
e^{\half\phi}~\sim{1\over\log u-({7\over8}+{g'(T_0)^2\over g(T_0)^2})
\log\log u}.
}
Let us compare this to perturbative Yang-Mills, where the coupling, up to
two loop order is,
\eqn\ptwoloop{
{g_{YM}}^2~\sim {1\over\log u-{b_2\over2 b_1^2}\log\log u}
}
and where $b_1$ and $b_2$ are the one and two loop contributions to the 
$\beta$-function\foot{A more careful calculation of the effective coupling
from the Wilson loop  shows that there are no additional
$\log\log u$ terms.  We use the convention that $b_1$ is negative and 
$b_2$ is positive.}.
For an $SU(N)$ gauge theory with six adjoint scalars, the ratio is
$b_2/2b_1^2=3/16$.  Hence we find that the sign of $b_2/2b_1^2$ coming from
dual gravity is model independent, and that the minimum value of
the ratio, and the one that comes closest to the perturbative result,
is $7/8$ when $g'(T_0)=0$.

\newsec{Connecting to the Infrared}

While it is satisfying that a reasonable {\it UV} solution exists, it is
not immediately clear what the {\it IR} behavior is like.  Numerical simulation
is very difficult with four coupled nonlinear differential equations,
so while useful, it does not immediately lead one to the answer.

To better explore the situation let us consider a greatly simplified
model where $f'(T_0)=g(T_0)=g'(T_0)=0$.  In this case the tachyon
is constant and the other three fields decouple from each other.  Obviously
the solution for the dilaton is $\phi=\a_0\rho+\phi_0$.  The other equations
of motion reduce to
\eqn\eqred{\eqalign{
\ddot\xi~+~{2Q^2\over f(T_0)}e^{-2\xi}~=~0\cr
\ddot\eta~+~8e^{-4\xi}~=~0.
}}
The $AdS_5\times S_5$ solution is $\xi=\log(\sqrt{2/f(T_0)}Q\rho$ and 
$\eta=\half\log(4\rho)$.  However, there can be other solutions to \eqred.  
Consider
the case where the solution asymptotically approaches the
$AdS_5\times S_5$ solution in the
limit $\rho\to0$.  Each second order differential equation should have
two integration constants.  Two of these are fixed by choosing the singularity
at $\rho=0$.  
These generalized solutions are given by
\eqn\solredgen{\eqalign{
\xi~=~\log(\sqrt{2/f(T_0)}~Q\rho)~+~\sum_{n=1}a_n\rho^{2n}\cr
\eta~=~\half\log(4\rho)~+~\sum_{n=1}b_n\rho^{2n},
}}
where $a_1$ and $b_1$ are free parameters and the higher $a_n$ and $b_n$
are determined by recursion relations.

We can also consider corrections to the {\it IR} ($\rho\to\infty$) 
$AdS_5\times S_5$ solution.  In this case the solutions have the form
\eqn\solredgen{\eqalign{
\xi~=~\log(\sqrt{2/f(T_0)}Q\rho)~+~\sum_{n=1}c_n\rho^{-n}\cr
\eta~=~\half\log(4\rho)~+~\sum_{n=1}d_n\rho^{-n},
}}
where $c_1$ and $d_1$ are free parameters and the higher $c_n$ and $d_n$
are computed via recursion relations.   It is straightforward to
show that a {\it UV} solution with
generic values for  $a_1$ and $b_1$ does not connect to the {\it IR} 
solutions in 
\solredgen\ for any  value of $c_1$ and $d_1$.   To see
this, note that $\xi$ and $\eta$ both satisfy an invariance equation,
\eqn\xietainv{
\half\dot\xi^2-{Q^2\over f(T_0)}e^{-2\xi}=C_1\qquad\qquad
-5\dot\eta^2+20e^{-4\eta}=C_2.
}
A quick calculation shows that for the {\it UV} solutions, $C_1=2a_1$ and 
$C_2=-10b_1$.  However, the {\it IR} solution has $C_1=C_2=0$ for any $c_1$
and $d_1$.  Hence, unless $a_1=b_1=0$, the {\it UV} solution does not connect
to this {\it IR} solution.  To see what solution it does connect to, note that
if $\xi>>1$, then the nonlinear term in the equation of motion is
small, likewise for the equation for $\eta$.  Hence for large $\xi$,
the leading order behavior for $\xi$ and $\eta$ is 
$\xi=\a_1\rho+\xi_0+{\rm O}(e^{-2\a_1\rho})$ and 
$\eta=\a_2\rho+\eta_0+{\rm O}(e^{-4\a_2\rho})$, with $\a_1,\a_2>0$.  Note
that these solutions are quite generic since they involve 4 integration
constants.
The constants of the motion are $C_1=\a_1^2/2$ and $C_2=-5\a_2^2$.
Hence we see that the {\it UV} solutions connect to the {\it IR} solutions with
$\a_1=2\sqrt{a_1}$ and $\a_2=\sqrt{2b_1}$.  If we include the $\a_0$
term from $\phi$, then the constraint condition in \constraint\ 
implies
\eqn\consimp{
\half\a_0^2~+~\half\a_1^2~-5~\a_2^2~=~0
}
 From the {\it UV} point of view, after fixing the coupling and
 the singularities at $\rho=0$, we are left with three integration constants.
The constraint reduces this to two.  Adjusting these integration constants
adjusts the values of $\a_i$, $i=0,1,2$, which also satisfy a constraint
relation.  Notice that one possible transformation  is the conformal
rescaling, where 
\eqn\conf{
\xi(\rho)\to\xi(t\rho)-\log(t)\qquad
\eta(\rho)\to\eta(t\rho)-\half\log(t)\qquad\phi(\rho)\to\phi(t\rho).
}
Obviously, under this transformation the $\a_i$ rescale to $\a_i\to t\a_i$,
and so the $\a_i$  can be set arbitrarily close to zero, the limiting
result being the usual {\it IR} fixed point of $AdS_5\times S_5$.
Aside from this conformal transformation, we are left with one independent
transformation to the integration constants.

Let us turn to the more difficult case where $f'(T_0)=g'(T_0)=0$, but
$g(T_0)\ne0$.  From the last line of \eqmot\ we see that $T$ is constant,
hence the problem is reduced to 3 coupled nonlinear differential equations.
As was argued in the last section \refs{\JM,\KTII}, 
there is a {\it UV} solution 
whose leading behavior is given by \asymp.  Adjusting $\xi$, $\eta$ and $\phi$
to all have singularities at $\rho=0$ takes care of three integration
constants.  Another integration constant is varied under the conformal
rescalings in \conf.  This leaves two integration constants.  These are
of a similar type as the previous example, although they are entangled
due to the coupling of the equations.  The leading correction
  for
 $\xi$ is $5a \rho^2$ while the leading
correction to $\eta$ is $a\rho^2$.  These also generate corrections
to $\phi$.  There is another integration constant appearing at order
$\rho^2$ which can be adjusted to satisfy the constraint in \constraint, 
leaving only one adjustable integration constant.

Notice that since the {\it UV} effective coupling is
$g_{eff}^2\sim1/(\log u)$, these subleading terms look like instanton
contributions.  In the perturbative regime, we should expect instanton
contributions to be powers of $u^{-8N/3}=\rho^{2N/3}$.  
Hence in the large $N$ limit,
no such terms should appear\foot{Unless of course $N=3$!}.  Since all
fields lie in the adjoint representation, it is possible that there
are fractional instantons.  However, these usually require fermion zero
modes which are conspicuously absent here.  

Another source for  $u^{-1}$ corrections are scalar masses.  These could be
bare masses or 
 masses generated by quantum corrections.  The ability to adjust the
integration constants on the type 0 gravity side could then correspond
to the freedom to adjust the bare masses on the field theory side\foot{
I thank I. Klebanov for a comment suggesting this.}.
The powers of $u$ could also just be an artifact of the strong coupling
expansion. In any event, since this theory is not conformally invariant, 
there does
not seem to be any {\it a priori} reason why they should be set to zero.
Hence, we will assume that these coefficients are fixed, but generic.

Let us now try to match to {\it IR} solutions.  One such solution was
discussed in \JM.  Consider the ansatz where $e^{-4\eta}$ is small
compared to the other terms in $V$.
This corresponds to a small curvature for the  $5$ sphere.  Dropping
this term, one can now
find an exact solution to the equations of
motion that satisfies the constraint.  The solution is
\eqn\lrhosol{
e^\phi~=~{C_0^2}\rho^{5/9}\qquad e^\eta~=~C_2\rho^{5/9}
\qquad e^{\xi}~=~{3Q\over\sqrt{2f(T_0)}}\rho,
}
with the relation
\eqn\conrel{
10(2f(T_0))^{1/4}C_2^5-9g(T_0)\sqrt{3Q}~C_0~=~0.
}
 Comparing all terms in $V$, one has
 $e^{\half\phi+\half\xi-5\eta}\sim \rho^{-2}$,
$e^{-2\xi}\sim \rho^{-2}$, but $e^{-4\eta}\sim \rho^{-20/9}$. Hence
this solution is valid for large $\rho$.  From \lrhosol, the 
coupling blows up as $\rho\to\infty$ and after substituting $\rho=1/u^4$
the metric is
\eqn\metsmrho{
ds^2~=~{10\over9 g(T_0)}\left(16{du^2\over u^2}+{C_2^5\sqrt{2f(T_0)}\over 3Q}
u^{8/9}
dx_{||}^2+C_2^5u^{-8/9}d\Omega_5^2\right).
}
$C_2$ remains as a leftover integration constant. 
 As in the {\it  UV}, the curvature in the {\it IR} is
 small if $g(T_0)<<1$.   However, there is no reason to expect
$g(T_0)$ not to be of order unity.  Hence, as in the {\it UV}, the string
corrections are of the same order as the classical
contribution.

While $e^\phi$ blows up for this particular {\it IR} solution, one can easily 
see that the effective coupling found from the heavy quark potential
reaches a finite value.
Defining a new variable $v$ such that 
\eqn\vrel{
v~=~{1\over9}\left({\sqrt{2f(T_0)}\over 3Q}\right)^{1/2}~u^{4/9},
}
 the metric in \metsmrho\ is then
\eqn\metv{
ds^2~=~{90\over g(T_0)}\left({dv^2\over v^2}~+~v^2dx_{||}^2~
+~{(C_2/3)^{10}\sqrt{2f(T_0)}\over Q^2v^2}d\Omega_5^2\right).
}
From this metric, we see that $R^2={90\over g(T_0)}$, and so the heavy quark
potential is \JM
\eqn\qpII{
V~\approx~-~{720\ \pi^3\over  \Gamma({1\over4})^4g(T_0)L}
\qquad\qquad L>>L_0.
}
Hence, we see that this particular solution is a conformal fixed point
in the {\it IR}.  

A possible cause of this behavior is that the $SU(N)$ gauge group has
been Higgsed to $U(1)^{n-1}$ by the adjoint scalars.  Naively, one expects
a repulsive force between the branes because of the extra R-R field.  In
the large $N$ limit, the branes could be pushed apart, but still maintain
the $SO(6)$ symmetry.  The effect would be to have the branes smeared out
over some region with spherical symmetry and finite size.  In the {\it UV},
well above the Higgs scale, the gauge group is unbroken and the coupling
runs.  In the infrared, the Wilson loop will probe down into this smeared
 region and  see the effect of the broken gauge symmetry, hence the 
quark potential behaves coulombically.

However, this ignores the role of the tachyon, which one might expect to
lead to attraction between the branes.  From the perturbative side this
seems reasonable, since quantum fluctuations would give masses to the 
scalar fields, removing the flat directions away from the unbroken
gauge theory.  There is a Coleman-Weinberg potential that is unstable
\Zarembo,
but this requires fine tuning to get rid of the scalar mass.  

Connecting this {\it IR} solution to the desired {\it UV} solution is 
problematic.
Just as in the uncoupled case, the generic {\it IR} behavior 
for $\rho\to\infty$ has the form 
\eqn\IRasymp{\eqalign{
\phi~&\approx~\a_0\rho~+~\phi_0\cr
\xi~&\approx~\a_1\rho~+~\xi_0\cr
\eta~&\approx~\a_2\rho~+~\eta_0.
}}
The constraints on $\a_i$ are 
\eqn\IRconstraint{
\a_i\ge 0,\qquad \half\a_0^2+ \half\a_1^2- 5\a_2^2=0,
}
where the latter constraint also insures that $5\a_1>\half\a_1+\half\a_2$,
thus insuring that all terms in the potential $V$ have an exponential falloff.
The {\it UV} solution is connected to the {\it IR} solution in \lrhosol\ 
by varying
integration constants such that the $\a_i$ in \IRasymp\ are set to zero.
 From the {\it IR} side
of things, we see that the $\a_i$ can all be set to zero by an infinite
rescaling.  But we do not want to do this since the rescaling will take the
{\it UV} coupling to zero at finite $\rho$.  Hence we only have one
free parameter
to work with.  Thus we see that it is not even
guaranteed that that there exists an $a_1$
and $b_1$ that will connect the {\it UV} solution to the {\it IR} conformal 
point.  In
fact, without some symmetry, it seems highly unlikely.  Even if it is
possible to set the $\a_i$ to zero, it is at best a horrendous fine tuning 
problem.

Let us thus accept that the {\it UV} solution attaches to an {\it IR} solution
with the asymptotic behavior in \IRasymp, with nonzero $\a_i$.  
In  section 5 we will
investigate the consequences of this.  Before doing this
let us round out this section
by discussing the most general case where $g'(T_0)\ne 0$.  In this
case $T$ will flow as $\rho$ is varied and thus we have 4 coupled
nonlinear differential equations.  If $g'(T_0)<0$ then  $T$ flows
toward zero as $\rho$ is increased.  One possible {\it IR} solution was
discussed in \KTII\ where $\xi$ and $\eta$ increase logarithmically,
$\phi$ increases as a log of a log, and $T$ relaxes to zero. The 
nice thing about this scenario is that the behavior of $g(T)$ and $f(T)$ 
are known as $T\to0$.
Unlike the previous case, there appear to be enough integration constants
in the {\it UV} to match solutions.    The generic solution is still of the 
form \IRasymp\ with the additional equation
\eqn\Tasymp{
T~\approx~\a_3\rho~+T_0.
}
The constraint equation restricts the $\a_i$ as
\eqn\IRTconstraint{
\half\a_0^2~+~\half\a_1^2~+~{1\over4}\a_3^2~-~5\a_2^2~=~0
}
The leading  correction to $T$ is 
$\sqrt{\rho}(T_+\rho^{\b_+}+T_-\rho^{\b_-})$, where 
\eqn\beq{
\b_\pm=\half(1~\pm~\sqrt{1-4f''(T_0)/f(T_0)}).
}
  If $f''(T_0)>f(T_0)/4$, which is the case for the function in \fT,
 then
we can write down an oscillatory solution with an amplitude and a 
phase shift.  Hence, naively anyway, we appear to have enough integration
constants to match solutions.  But matching to this particular {\it IR}
solution still requires
a tremendous amount of fine tuning.  

If $f''(T_0)\le f(T_0)/4$ then $\b_\pm$
would both be positive real and there would still be two integration constants
to adjust.  This behavior would also fix the problem of instabilities due
to the negative tachyon mass near the {\it UV} fixed point.  At this point,
the space is locally $AdS_5$ and so the condition for stability is that
the tachyon mass satisfy $m^2\ge-4$ \refs{\BF,\MeTo,\Witten}.  This is
precisely the condition that $f''(T_0)\le f(T_0)/4$.  If this condition
were satisfied then at the ${\it UV}$ fixed point the tachyon would couple
to a relevant operator.

More generic is the case where $T$ relaxes to zero or some other
constant $T_{IR}$ while
the other variables have the asymptotic behavior in \IRasymp.  In this
case, $T-T_{IR}$ falls off exponentially as $\rho\to\infty$.  We can see
from the equations of motion for $T$ that the second derivative starts out
positive, but  becomes negative at some point before $T=0$. Hence, it 
should be possible to find a solution where $T$ starts upward but then
relaxes to some finite point, while all the other fields grow linearly in
$\rho$. 

Finally, let us consider  $g'(T_0)>0$.  Now $T$ flows 
away from $T=0$ and  there appear to be no possible {\it IR} solutions except
for those of \IRasymp\ with $\a_i>0$ and \Tasymp\  with $\a_3<0$.

\newsec{Possible Validity for a Class of Infrared Solutions}

In order to trust the solutions with the asymptotic behavior in \IRasymp\ and
\Tasymp, it is necessary to check that the higher $\a'$ corrections are
small.  At first glance, this would appear to be a miserable failure.  The
curvature in the Einstein metric
$ds_E^2=e^{-\phi/2}ds^2$, where $ds^2$ is the metric in \typezm, approaches
\eqn\curv{
R~\approx~\left(\half{\a_1}^2-5{\a_2}^2\right)e^{(5\a_2-\a_1/2)\rho}.
}
This blows up as $\rho\to\infty$ because of the constraint in \IRTconstraint.

The leading string corrections are of the form ${\a'}^3 e^{-3\phi/2}W$,
where $W$ is a combination of four contracted Riemann tensors.  Hence the naive
dependence for the ${\a'}^3$ terms using the solutions in \IRasymp\ is
\eqn\Rfour{
e^{-3\phi/2}W~\sim~e^{(20\a_2-2\a_1-3\a_0/2)\rho}.
}
Using the conditions in \IRTconstraint, one immediately sees that this
term grows faster than $R$, and hence the classical approximation breaks
down.

However, $W$ has a field redefinition ambiguity and can be written
strictly in terms of the Weyl tensor $C_{\mu\nu\la\d}$.  For the
Einstein metric, the nonzero components of the Weyl tensor have the
form
\eqn\Weyl{
C_{\mu\nu\mu\nu}~=~g_{\mu\mu}g_{\nu\nu}(A_1{\cal F}~+~A_2
{\cal G})
e^{5\eta-\xi/2},}
where $A_1$ and $A_2$  are constants (different components have different
constants), and ${\cal F}$ and ${\cal G}$ are defined as
\eqn\FG{
{\cal F}~=~2(\dot\xi-\dot\eta)\dot\eta~+~\ddot\xi~-~\ddot\eta,\qquad\qquad
{\cal G}~=~\dot\xi^2~-~\dot\eta^2~+~\ddot\xi~-~\ddot\eta~-~4e^{-4\eta}.
}

It is clear that for a general solution in $\IRasymp$, both ${\cal F}$ and
${\cal G}$ are of order unity, and so $e^{-3/2\phi}W$  blows up as in
\Rfour.  However, if $\a_2=\a_1$, then both terms are suppressed.  We
actually should have anticipated this, since now the Einstein metric
approaches that of the product space $R_+\times M_4\times S_5$, up to
a conformal transformation.  Recall that the metric for the 
3(4) dimensional Witten model approaches the metric for the product space
$R_5\times S_5$ ($R_6\times S_4$) near the black hole horizon.

However, simply setting $\a_2=\a_1$ is not sufficient to insure that
$e^{-3\phi/2}W$  blows up slower than $R$.  From the equations
of motion in \eqmot, we see that $\xi$ has corrections of
order $e^{-2\a_1\rho}$ and $\eta$ has corrections of order 
$e^{(\a_0+\a_1-10\a_2)\rho/2}$.  Letting $\a_2=\a_1$, we find  in
general that ${\cal F}\sim e^{-2\a_1\rho}$ and  ${\cal G}\sim e^{-2\a_1\rho}$,
and thus
\eqn\Wasymp{
e^{-3\phi/2}W~\sim~e^{(10\a_1-3\a_0/2)\rho}.
}
While the behavior has been softened, it is still not enough to prevent the
${\a'}^3$ correction from dominating $R$, since the maximum value for $\a_0$
consistent with \IRTconstraint\ is $\a_0=3\a_1$.

However, there are solutions where $W$ grows even slower than
\Wasymp.  If
$T$ approaches a constant $T_{IR}$ as $\rho\to\infty$, then
$\a_3=0$ and the asymptotic
solution has the relations
\eqn\goodasymp{
\a_0~=~3\a_1,\qquad\qquad\a_2=\a_1.
}
For this case, $\xi$ has corrections of order 
$e^{-2\a_1\rho}$ and $e^{-3\a_1\rho}$ 
while $\eta$ has corrections of order $e^{-3\a_1\rho}$.  
It is a straightforward excercise
to check that these corrections cancel in {\it both} ${\cal F}$ and ${\cal G}$
and therefore 
\eqn\goodW{
e^{-3\phi/2}W~\sim~e^{-5\a_1\rho/2}.
}
Hence, not only does the term grow slower than $R$, it actually falls off
exponentially!

We can also check the behavior for other types of terms.  One class of
terms has the form \KTII
\eqn\tI{
e^{\phi/2}\left(e^{-\phi/2}C\right)^n~\sim~e^{-(2n-3)\a_1\rho/2},
}
falling off even faster than the first string correction.  There are also
terms of the form 
\eqn\tII{
e^{-(n-1)\phi/2}(\nabla\phi\nabla\phi)^n~\sim~e^{-3(n-1)\a_1\phi/2},
}
which again falls off rapidly.  Terms involving derivatives of the tachyon
field also have exponential suppression since $T$ approaches a constant
value with exponential falloff.  The behavior of all these corrections
 leads us to speculate that there is a background that is
an exact solution of the
$\sigma$ model which has the asymptotic behavior defined by
\IRasymp\ and \goodasymp.


There is some fine tuning involved in connecting the desired {\it UV} solution
to the {\it IR} solutions in \goodasymp.  We need to tune such that $\a_1=\a_2$
and $\a_3=0$.  If $g'(T_0)=0$ then $\a_3=0$ is automatically satisfied.
In either case there are enough available  integration constants on the 
{\it UV} side which  can be adjusted
 so that \goodasymp\ is satisfied.  Hence, we believe
that there is a {\it UV} solution that connects to the {\it IR} solutions in 
\IRasymp\ and \goodasymp.

\newsec{Confinement in the Infrared}

In this section, we argue that the {\it IR} solutions of \IRasymp\ and \Tasymp\
lead to a linear quark potential,  magnetic screening and a discrete spectrum.
The linear quark potential requires the additional condition $\a_1\ge\a_0$.
Happily, the solution in \goodasymp\ satisfies this condition.  It is
possible that other solutions besides \goodasymp\ are valid, perhaps
there is enough ambiguity in the effective action to make their
$\a'$ corrections small as well.   To allow for this, we 
will discuss the general case in this section.

\subsec{The Quark Potential}

The quark potential is computed using the Wilson loop calculation of
\refs{\RY,\JMaldb}.  The Nambu-Goto action for a string in a curved
background is given by
\eqn\NG{
S~=~{1\over2\pi}\int d\s\d\tau \sqrt{det(G_{MN}\partial_\a X^M\partial_\b
X^N)}.
}
The quark potential is computed from the rectangular Wilson loop, with
two sides along the time direction and the other two along one of the
spatial directions.  Using the metric in \typezm, we find that the
energy for the quark-antiquark pair is
\eqn\qpot{
E~=~{1\over2\pi}\int dx\sqrt{e^{\phi-5\eta}\Big({\partial\rho\over\partial x}
\Big)^2~+~e^{\phi-\xi}}.
}
Changing variables such that $du=-e^{(\phi-5\eta)/2}d\rho$, we have
\eqn\qpotII{
E~=~{1\over2\pi}\int dx\sqrt{\Big({\partial u\over\partial x}
\Big)^2~+~f(u)/Q},
}
where $f(u)$  has the asymptotic behavior
\eqn\fuasymp{\eqalign{
f(u)~\sim~ {u^4\over (\log u)^2}\qquad\qquad u>>1\cr
f(u)~\sim~u^\g\qquad\g={2(\a_1-\a_0)\over5\a_2-\a_0}\qquad\qquad u<<1.
}}
Assuming that all $\a_i\ge0$, the constraint equation \IRTconstraint\
implies that the exponent for the small $u$ behavior satisfies
$\g<2$.
If $\a_1>\a_0$, then $0<\g<2$, and thus for large enough $x$, the minimum
energy configuration consists of a string starting at $u=\infty$ and
going straight down to the origin at $u=0$, traveling a distance $x$ along
$u=0$, and then going back out to $u=\infty$.  Since $f(u)=0$ at $u=0$,
there is no cost in energy to separate the quarks further, thus this
would correspond to electric screening \refs{\RTY,\BISY}.

If instead $\a_1=\a_0$, then $u^\g$ is no longer zero at the origin.  In
this case, for large enough $x$ the string will approach the origin
and the quark potential becomes linear.   If $\a_1<a_0$, then $u^\g$ blows
up at the origin.  In this case there will be some nonzero $u_0$ such
that $f(u_0)$ is a minimum.  For large enough $x$ the string will 
approach this minimum value. Since $f(u)$ is clearly positive for
nonzero $u$, this too will lead to a linear quark potential.  Assuming that
the $\a_i$ are all of order 1, then the string tension will be of
order $1/\sqrt{Q}$ in the units used here.  

If $\a_1<\a_0$,  then the coupling
seen by the string is  in some sense bounded in the infrared. 
For
large quark separation, the  minimum energy configuration has $u>u_0$,
thus the coupling approaches $e^{\phi_0+\a_1u_0}$.  This is not to say
that the behavior of the metric and coupling for $u<u_0$ will not affect
other physical results, as we will see in the following subsections. 

\subsec{Magnetic Screening}

If there is confinement, then there should be a corresponding screening
of magnetic charge.  The computation is along the lines of \GO.  In particular,
we wish to compute the potential between a monopole anti-monopole pair.
This is accomplished by calculating the rectangular Wilson loop for
a D string.  In type 0 theory, since there are two types of R-R Kalb-Ramond
fields, there are also two types of D strings.  However, only one type
of D string can end on the electric D3 brane.  It is this string that
describes the Wilson loop for a monopole.

The calculation is almost identical to the Wilson loop calculation for
the fundamental string.  The only difference is that the integrand in
\NG\ is multiplied by a factor of $e^{-\phi}$.  Hence, the potential is
given by
\eqn\mpot{
E~=~{1\over2\pi}\int dx\sqrt{e^{-\phi-5\eta}\Big({\partial\rho\over\partial x}
\Big)^2~+~e^{-\phi-\xi}}.
}
This time changing variables such that $du=e^{(-\phi-5\eta)/2}d\rho$, we have
\eqn\qpotII{
E~=~{1\over2\pi}\int dx\sqrt{\Big({\partial u\over\partial x}
\Big)^2~+~f(u)/Q},
}
where now $f(u)$  has the asymptotic behavior
\eqn\fuasymp{\eqalign{
f(u)~\sim~ {u^4 (\log u)^2}\qquad\qquad u>>1\cr
f(u)~\sim~u^\g\qquad\g={2(\a_1+\a_0)\over5\a_2+\a_0}\qquad\qquad u<<1.
}}
Since all $\a_i\ge0$ and $5\a_2>a_1$, then $0<\g<2$.  Hence, the monopole
and anti-monopole will screen at large distances

\subsec{WKB Estimates for the Glueball Spectrum}

The correlator $\langle \Tr(F^2(x))\Tr(F^2(0))\rangle$  is related to
scalar Green's functions in the bulk \refs{\GKP,\Witten}.  A discrete
spectrum for the bulk particles would correspond to a discrete spectrum
for the scalar glueballs \Wittherm. 
Since we are dealing with string tensions of order 1, the green's
functions of massive string states could be important.  More to the point, the
tachyon couples to $\Tr(F^2)$ \refs{\KTII,\Gar}.  If the tachyon is a bad
tachyon, with plane wave normalizable states in the bulk, then it
cannot have a discrete spectrum.  Even if it is well behaved with
$f''(T_0)<f(T_0)/4$, one might find that the lowest lying state in
the spectrum is tachyonic.  

The linear equations one derives for
the small fluctuations are hopelessly entangled through the tachyon mass
term.  As a first approximation, one can ignore the cross terms and simply
consider the diagonal equations. Let us make the usual ansatz and
assume that the $0^{++}$ glueball spectrum
is determined by the spectrum for small fluctuations of the dilaton
equation of motion. 
Consider dilaton solutions of the form $\phi=\psi(\rho)e^{ik\cdot x}$, where
$k^2=-M^2$.  Then the dilaton equation of motion reduces to
\eqn\dileqm{
{\partial\over\partial\rho}\sqrt{g}e^{-2\phi}g^{\rho\rho}
{\partial\over\partial\rho}\psi~+~M^2\sqrt{g}e^{-2\phi}g^{xx}\psi~+~
{1\over4} g(T)\sqrt{g}\psi~=~0,
}
where $\phi$ in \dileqm\ refers to the background dilaton field.  Substituting
the metric in \typezm\ into \dileqm\ and using the equation of motion for
the background dilaton field, one finds
\eqn\dileqmr{
{\partial^2\over\partial\rho^2}\psi~+~M^2e^{\xi-5\eta}\psi~-~
{1\over2} \phi''\psi~=~0.
}
Defining $u$ as $e^{\xi-5\eta}=Qu^6\sqrt{2/f(T_0)}/32$, the equation of
motion becomes
\eqn\dileqmu{
{\partial\over\partial u}F(u){\partial\over\partial u}\psi~+~
M^2QH(u)\psi~+~G(u)\psi~=~0,
}
where $F(u)$, $H(u)$ and $G(u)$ have the asymptotic behavior 
\eqn\FHasymp{\eqalign{
F(u)&~\approx~u^5\qquad\qquad  H(u)~\sim~ u\qquad\qquad
G(u)~\approx~{4u^3\over\log u}  \qquad\qquad u>>1\cr
F(u)&~\sim~u\qquad\qquad  H(u)~\sim~ u^5\qquad\qquad G(u)~\sim~ u^{7/2}
\qquad\qquad u<<1.
}}
Note that the substitution for $u$ requires that $5\a_2>\a_1$ in order
that $u\to0$ as $\rho\to\infty$.  Luckily, this inequality is guaranteed by
\IRTconstraint.

Using the arguments in \Wittherm, one can argue that the spectrum is
discrete.  
In fact, we can say more.  Since we know the asymptotic behavior,
one can do the WKB calculation described in \JM.  The asymptotic behavior
is  similar  to that of
 the three dimensional Witten model \Wittherm, although there will be
log corrections because of the {\it UV} behavior of $G(u)$.  The $G(u)$ 
term does not affect the {\it IR} turning point to next to leading order.
Borrowing the results in \JM\ and inserting the $G(u)$ correction, 
we find that the WKB $0^{++}$ glueball spectrum is
\eqn\WKBgb{
M^2~=~ {4C\sqrt{2f(T_0)}\over Q}~ m\left(m+1-{2\over \log [C m(m+1)]}\right)
~+~{\rm O}\left(m\over \log^{2}m\right)\qquad\qquad m\ge 1,
}
where $C$ is a constant of order unity.  
Corrections to \WKBgb\ coming from the inclusion of cross terms are
 of order $m/\log^2m$.
The  log factors in \WKBgb\ 
have the effect of dragging down the glueball masses, which is clearly 
caused by the running of the coupling.  This is especially
true for the lightest state.  If $C=1$ and one assumed that \WKBgb\ were
exact, then one would find that the lightest state is actually tachyonic.
At this point we are unable to prove that a tachyonic state does not appear.
  We can say that if a tachyonic
state were to exist for the dilaton equation in \dileqm, 
then a minimum requirement is that there is some region
where $\rho^2\phi''<-1/2$.  One can see that this is never true if we
only consider the leading order term in \asymp\ for $\phi$.  Numerical
analysis seems to show that it is not true if the higher order terms are
also included. Once we are safely in the infrared, then $\phi''$ will be 
exponentially suppressed and so $\rho^2\phi''>-1/2$ in this region as well.
However, inclusion of the cross terms might drop the masses enough to leave
a tachyonic glueball state.  Hopefully this does not occur.

Since $Q\sim N$, we see that the glueball masses are a factor of $N^{1/4}$
smaller than the square root of the string tension.  This is not as severe
as the 4 dimensional Witten model, where the masses are a factor of $N^{1/2}$
smaller \GrOl.  We can also compare the ratio of the first excited glueball
to the lowest mass state.  Since our knowledge of $C$ is limited, we can
only put a lower bound on the mass ratios, which is reached when $C\to\infty$.
Plugging in numbers, we find that this ratio satisfies
$m_{++}^*/m_{++}>\sqrt{3}\approx 1.73$.  
The WKB approximation for this ratio in the 4 
dimensional Witten
model is $\sqrt{8/3}\approx1.63$ \JM.  The $SU(3)$ lattice result 
is $1.77\pm.14$ \Teper.

We can also make WKB approximations for the glueball states with nontrivial
$SO(6)$ quantum numbers \ORT.  If we write the dilaton field as
$\phi=\psi(\rho)e^{ik\cdot x} Y_l(\Omega_5)$, then the dilaton equation
of motion reduces to
\eqn\dileqmY{
{\partial^2\over\partial\rho^2}\psi~+~M^2e^{\xi-5\eta}\psi~-
l(l+4)e^{-4\eta}\psi-\half\phi''\psi~=~0.
}
If $\a_1\ge\a_2$, which is the case for \goodasymp, 
then the $M^2$ term will dominate the $\ell(\ell+4)$
term as $u\to 0$. 
In this case, the WKB calculation is a 
straightforward generalization of the $l=0$ result and  is
\eqn\WKBgbl{
M^2~=~{4C\sqrt{2f(T_0)}\over Q}~m\left(m+1+l-{2\over \log [C m(m+1)]}\right)
~+~{\rm O}\left(m\over \log^{2}m\right)\qquad\qquad m\ge 1.
}
If $\a_1<\a_2$, then the asymptotic behavior changes  near $u=0$, but this
effect only leads to corrections of order 1 in $m$. 

We complete this subsection by computing the WKB spectrum for the $0^{-+}$
glueballs.  In the Witten model, this was done by studying the
wave function for the
5 dimensional dual of the NS-NS two form field, which is a vector
\refs{\COOT,\HashOz}.  
The component along the Euclidean time direction is a scalar in the
four dimensional theory and is odd under parity.  In the case that we
are studying, there is no euclidean circle, so we need to consider
instead  the wave function for a scalar field.  The scalar field is
the IIB axion which is an
R-R scalar.  Actually, there are two scalars, but only one
couples to $\Tr(F\widetilde F)$ on the electric D3 branes.  The other scalar
couples to $\Tr(F\widetilde F)$ on the magnetic D3 branes.

Since the axion comes from the R-R sector, it does not couple to the 
dilaton\foot{The axion does couple to the tachyon \KT, but this does not 
affect the WKB approximation to next to leading order.}.
Hence, the wave function has the form
\eqn\axioneqm{
e^{-2\phi}{\partial\over\partial\rho} e^{2\phi}{\partial\over\partial\rho}
 a(\rho)~+~M^2 e^{\xi-5\eta}~=~0.
}
Defining $a(\rho)=e^{-\phi}\psi(\rho)$, \axioneqm\ can be rewritten as
\eqn\axionII{
{\partial^2\over\partial\rho^2}\psi~+~M^2e^{\xi-5\eta}\psi~-(\phi''+{\phi'}^2)
\psi~=~0.
}
As was the case for the dilaton,
 the $\phi$ terms do  affect the next to leading
order contribution
of the {\it UV} turning point.  
The ${\phi'}^2$ term also changes the next to leading order contribution of 
the {\it IR} turning point.  
Using the analysis in \JM, one finds that
the WKB mass expression for the $0^{-+}$ states are
\eqn\WKBax{
M^2~=~{4C\sqrt{2f(T_0)}\over Q}~m\left(m+1+{2\a_0\over 5\a_2-\a_1}
-{4\over \log [C m(m+1)]}\right)~+~{\rm O}\left(m\over \log^{2}m\right)
\ \  m\ge 1
}
where $C$ is the same constant as in \WKBgb.  Unlike the $0^{++}$ states, but 
like the $0^{-+}$ states in the Russo generalization of the Witten model
\refs{\Russo,\CORT,\CRST},
there is model dependence in these glueball masses.  
Using the solutions 
in \goodasymp\ we find as a lower bound for the mass ratio
$m_{-+}/m_{++}>\sqrt{7}/2\approx 1.32$.
The most recent lattice results have $m_{-+}/m_{++}=1.78\pm.24$ for $SU(2)$
\Teper\
and $m_{-+}/m_{++}=1.50\pm.04$ for $SU(3)$ \MP

\newsec{Discussion}

One of the main differences between the results found here and 
 results found for the Witten
model is that the type 0 model has a QCD
 scale set by dimensional transmutation.  Hence, the glueball mass is
directly related to the string scale.  We have also found differences for the
mass ratios.  The principle reasons for these differences
 is that the type 0
model is four, not five dimensional in the {\it UV},
 and because the type 0 model has a running
coupling constant.  The $N$ dependence in the
mass to string scale ratio is also affected by the differences in
dimension.  The Witten model has a factor of $N^{-1/2}$,
while the type 0 model has a factor of $N^{-1/4}$

The existence of the  confining solutions  does not contradict other 
results for {\it IR} fixed points.  The fine tuning involved to reach such 
solutions
could correspond to tuning mass parameters on the field theory side, so
that for instance, the six scalars are massless after including their
quantum corrections.  In this case, the field theory could have
the {\it IR} fixed point described in \KTII.  The confining solutions,
which do not need to be fine tuned, have massive scalars, and so are in
the same universality class as pure $SU(N)$.\foot{I thank I. Klebanov for
suggesting this point.}

We have seen that while confinement in the infrared is generic, there exists
solutions where many of the string corrections vanish exponentially.
One such solution has a constant tachyon expectation value.  In some sense,
this is closer to the spirit of Polyakov's noncritical string for QCD,
where the nonzero tachyon expectation value, ``peacefully condensing in the
bulk'', plays the role of a nonzero
central charge \refs{\Cave,\AG}.  On the other hand, the five form flux
through $S_5$ clearly plays an important part here; without it we would 
not have the log scaling in the {\it UV}.

\goodbreak
\vskip2.cm\centerline{\bf Acknowledgements}
\noindent

I would like to thank Igor Klebanov
for very helpful discussions and comments on the manuscript.  
I  would also like to thank John
Schwarz and  the Caltech Theory Group for hospitality and support
during the early stages of this work.
This research was partially supported by
  NSF  Grant DMS-9627351.

\goodbreak

\listrefs
\end